\documentclass[twocolumn,showpacs,prl,10pt]{revtex4}
\usepackage[english]{babel}
\usepackage{amsmath,amssymb}
\usepackage{times}
\usepackage{epsfig}

\begin{document} 
%\draft
\title{Reply to the Comments by B. Farid and A.M. Tsvelik,
  arXiv:0909.2886v1, and B. Farid, arXiv:0909.2887v1}

\author{J. Kokalj$^{1}$ and P. Prelov\v sek$^{1,2}$}
\affiliation{$^1$J.\ Stefan Institute, SI-1000 Ljubljana, Slovenia}
\affiliation{$^2$ Faculty of Mathematics and Physics, University of
Ljubljana, SI-1000 Ljubljana, Slovenia}

\date{\today}
                   
\begin{abstract}
In the recent preprints \cite{farid1,farid2} the authors claim that
conclusions on the violation of the Luttinger sum rule (LSR) as put
forward in published papers by the present authors \cite{kok1,kok2}
are false or at least not conclusive enough. It concerns the validity
of the LSR within the Mott-Hubbard insulator as manifested in the 1D
model of spinless fermions \cite{kok1}, and the Hubbard model on a
triangular lattice \cite{kok2}. We show below that arguments
against the presented analysis are rather easily rejected while
still the general question of the validity and breakdown of the LSR
remains to be answered.  
\end{abstract}

\pacs{71.10.-w, 71.27.+a}
\maketitle

The Luttinger theorem is clearly an essential pillar on which
the Fermi liquid theory for metals rests. Recently the extensions of
the Luttinger sum rule (LSR) to insulators has been proposed as well
as the possibility of the breakdown of the LSR has been considered by
several authors. In our recent paper \cite{kok1} we have shown that
for the simple one-dimensional (1D) tight-binding model of interacting
spinless fermions $t$-$V$ the LSR can be broken within the
Mott-Hubbard insulating phase provided that the electron-hole symmetry
is broken via the introduction on the next-nearest-neighbor (n.n.n.)
hopping $t'$.  In a long and exhaustive Comment B.Farid and
A.M. Tsvelik \cite{farid1}, which we are able to follow only
partially, give several arguments why our analysis is not
correct and the LSR remains in fact valid if interpreted properly.
Below we go through the arguments as they follow:

1) The proper definition of the chemical potential $\mu$ as the
zero-$T$ limit of $\mu(T\to 0)$ is essential
for the proof of the LSR and its validity whereby we agree with
the Comment \cite{farid1}. We are fully aware of this requirement and
follow it everywhere  therefore do not understand why we are  
cited on p.2 as ``not to have appreciated this fundamental aspect ``.

2) We agree with the Comment that LSR should apply also to finite-size
systems which facilitates the discussion of the validity as well as
possible breakdown of the LSR.

3) In Comment, p.3., it is claimed that our numerical results do not
really show the violation of the LSR and that the whole argument is
(essentially) based on delicate scaling of finite-size results.
This is not true, since even at fixed large sizes we get direct
evidence for violation. E.g., for $V/t=8$ and results obtained on
$N=26$ sites on Fig.2 it is evident for discrete $k_1=7 \pi/13> \pi/2$
$G(k_1,\omega=0) \sim 0$ being quite evident deviation from LSR.
Even more, results for $N=30$ (not presented in the paper)
show for the same parameter $V/t=8$ already for $k_2=8 \pi /15$
$G(k_2,\omega=0) > 0$, directly violating LSR on a finite-size
system. We should also stress that the deviations from LSR as evident in
Fig.~5 are quite substantial for $t'/t=0.4$ reaching up to $10 \%$
well beyond any numerical uncertainty.

4) In the Comment the authors present a toy-model example which
should indicate possible inaccuracies or failure of finite-size
scaling. It is, however, evident that the form as shown in their Fig.1
is quite pathological with $G(k,\mu)$ having the zero at $k=\pi/2$ but
at the same time a near divergency or pole at $|k| <\pi/2$. This pole in
fact fully appears for their parameter $A=1.97$ very close to cases
presented in the text. So very strong $k$ variation is understandably
the origin of delicate size dependence and poor scaling (in fact their
scaling works still very well for $A=1.6$ as presented in their Fig.1
and gives $k=\pi/2$ to high accuracy). On the other hand, our form of
$G(k,\mu)$ vs. $k$ as shown in Fig.~2 is very modest and even
monotonous, at least for the case with a pronounced gap, i.e.  $V/t
\geq 4$ as analysed also in Figs.~3-5.

5) The possibility of a ordered state of charge-density-wave (CDW)
type as mentioned on p.6 is clearly of relevance. However, the
ordered state can appear only in the limit $N \to \infty$ while in all
finite systems the ground state still has not broken translational
invariance with $k=0$. Since our results show substantial deviations 
of the LSR they appear already much before the limit $N\to \infty$.
Also it should be pointed out that the long-range order does not
necessarily invalidate the LSR but merely needs a redefinition of the
latter as correctly pointed out in the Comment.

In the other Comment \cite{farid2} the author claims that the results
for the Hubbard model on a triangular lattice indicating the violation
of the LSR for $U/t \gg 1$ \cite{kok2} are errorneous. This is somewhat
surprising since the author follows in detail our original steps in
the analytical expansion for $t/U \ll 1$ and does not find any
inconsistencies in our analysis. Only on his p.3 (after Eq.~(28))
he claims that there is a 'fallacy in the reasoning' although it is hard
to establish the exact objection:

1) It seems that the author starts from the assumption in the limit of
$t/U \to 0$ the LSR should hold ? However, the latter limit is far
from trivial, and there is no guarantee for LSR, or even more the LSR
appears to fail rather evidently as shown in our paper \cite{kok2}.
Moreover, certain caveats (as pointed out in our analysis) are also
necessary for strictly $U/t \to \infty$ since the ground state can
become of different symmetry (large $S$), at least for small systems.

2) The author \cite{farid2} seems to object to the expansion 
in the deviation $\tilde \mu$. It is true that the latter
quantity we are not able to derive analytically (in contrast
to $G({\bf k},\mu_0)$) still numerical results for different
sizes (in Fig.~1) clearly give a consistent scaling.

3) Author in his Comment \cite{farid2} estimates the discrepancy
between our first order approximation (in $t_{ij}/U$ expansion) of the
Luttinger wave vector 
$k_L^{(1)}$ and exact Luttinger wave vector $k_L$  with  the
order of $\tilde \mu$,
 \begin{equation}
k_L^{(1)}-k_L\sim \tilde \mu=O(t_{ij}).
\end{equation}
$k_L$ (following from $G(k_L,\mu)=0$) would hypothetically be obtained
with taking into consideration all orders of large $U$ expansion, but would
not neccessarilly satisfy the LSR.
This leads him to the conclusion that \cite{farid2}
\begin{equation}
\frac{G^{(0,2)} ({\bf k}_L^{(1)};U/2) }
  {G^{(1,0)} ({\bf k}_L;\mu) } =O(\frac{1}{\tilde \mu}).
\end{equation}
This two estimates seems rather ambiguous considering the dimensional
analysis and we show that within the $t_{ij}/U$ expansion
they are false. Both, numerator and denominator can within the
strong coupling expansion be easily estimated. With the use of 
Eqs. 4, 7 and 14 in reference \cite{kok2} one obtains for the numerator
$G^{(0,2)} ({\bf k}_L^{(1)};U/2) =O(t_{ij}/U^4)$ and with the use of
Eq. 16 in reference \cite{kok2} one readily sees that $G^{(1,0)} ({\bf
  k}_L;\mu)= O(t_{ij}/U^2)$.  All together gives the true estimate for
the discrepancy
\begin{equation}
k_L^{(1)}-k_L\sim O(\frac{t_{ij}^2}{U^2}),
\end{equation}
which can actually be quantitatively estimated using quantities
presented in \cite{kok2}  with much less than 1\%. 

In conclusion, considering in detail the Comments \cite{farid1,farid2}
we did not find any criticism of our analysis \cite{kok1,kok2} to be
founded. Still we agree with the author that the LSR and its possible
violation should be considered very carefully and that this subject
and its consequences are of fundamental importance for the solid state
physics and correlated systems in particular.

\end{document}